\documentclass[11pt]{article}

\usepackage[letterpaper, margin=1.25in,tmargin=1.25in,bmargin=1.25in]{geometry}

\usepackage{cite}

\usepackage{color}
\usepackage{amsmath}
\usepackage{graphicx}
\usepackage[noend]{algpseudocode}
\usepackage{algorithm}
\makeatletter
\def\BState{\State\hskip-\ALG@thistlm}
\makeatother

\usepackage[nolist]{acronym}

\usepackage{authblk}

\usepackage{color}

\begin{document}

\begin{acronym}
\acro{DASH}{Dynamic  Adaptive Streaming over HTTP}  
\acro{QoE}{Quality of Experience}  
\acro{NUM}{network utility maximization} 
\end{acronym}



\title{Price-based Controller for Quality-Fair HTTP Adaptive Streaming (Extended Version)}

\author[1]{Stefano D'Aronco}
\author[2]{Laura Toni}
\author[1]{Pascal Frossard}
\affil[1]{LTS4\\ Ecole Polytechnique F\'{e}d\'{e}rale de Lausanne (EPFL)}
\affil[2]{Electrical and Electronic Departement\\ University College London (UCL)}

\date{}

\maketitle
\begin{abstract}
HTTP adaptive streaming (HAS) has become the universal technology for video streaming over the Internet. Many HAS system designs aim at sharing the network bandwidth in a rate-fair manner. However, rate fairness is in general not equivalent to quality fairness as different video sequences might have different characteristics and resource requirements. In this work, we focus on this limitation and propose a novel controller for HAS clients that is able to reach quality fairness while  preserving the main characteristics of HAS systems and with a limited support from the network devices. In particular, we adopt a price-based mechanism in order to build a controller that maximizes the aggregate video quality for a set of HAS clients that share a common bottleneck.
When network resources are scarce, the clients with simple video sequences reduce the requested bitrate in favor of users that subscribe to more complex video sequences, leading to a more efficient network usage.
The proposed controller has been implemented in a network simulator, and the simulation results demonstrate its ability to share the available bandwidth among the HAS users in a quality-fair manner.
\end{abstract}




\section{Introduction}
\begin{figure}
\centering
\includegraphics[scale=0.115]{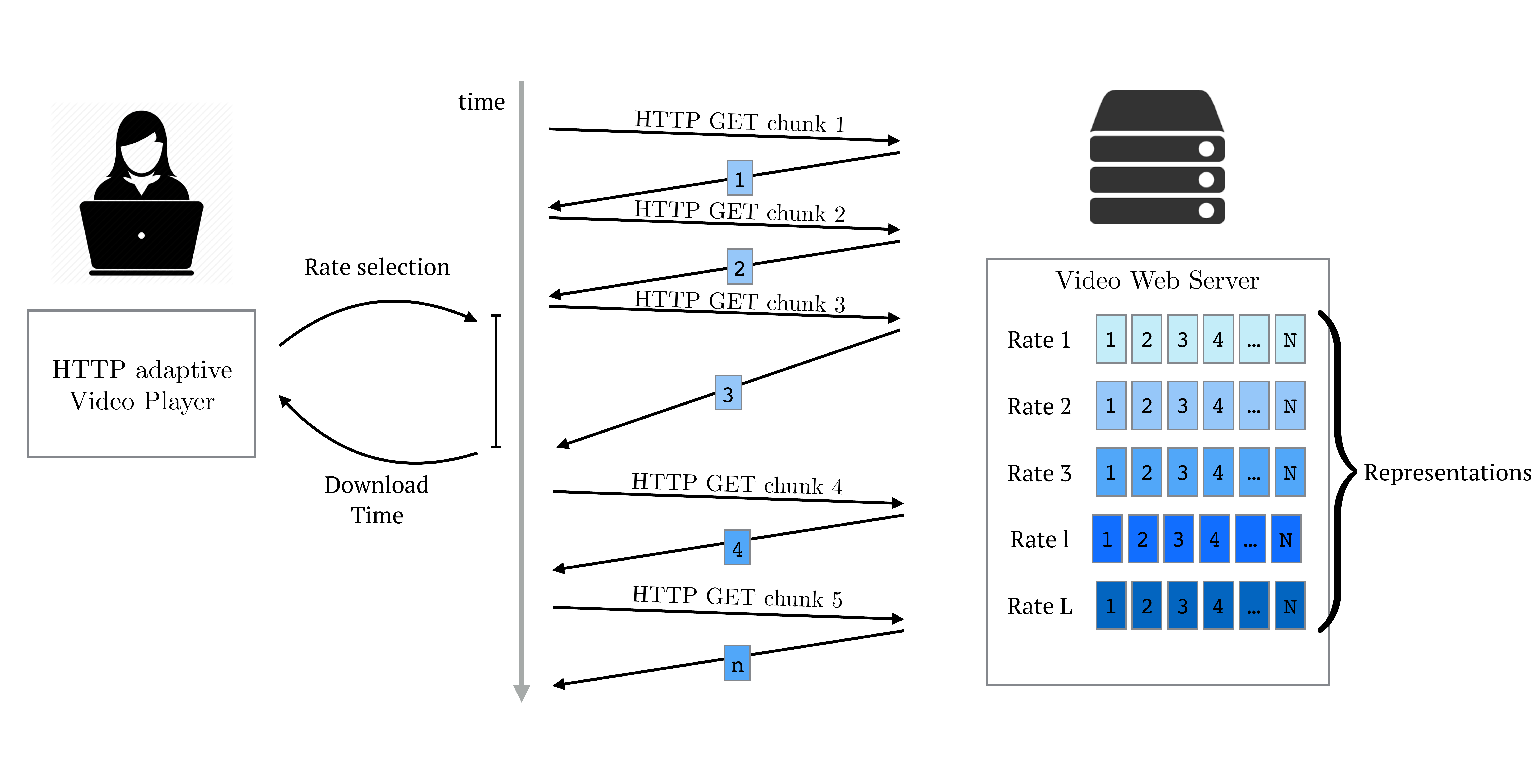}
\caption{General HAS system architecture.}
\label{fig:system}
\end{figure}

HTTP adaptive streaming (HAS) has become the universal client-driven streaming solution for video distribution over the Internet, an example of this paradigm is given by the Dynamic Adaptive Streaming over HTTP~\cite{stockhammer} (DASH) standard. In HAS, as it is shown in Fig.~\ref{fig:system}, the video content is available at the main server in different coded versions, namely representations,   each one with a given bitrate and resolution.  The representations are subdivided into chunks of few seconds typically, which are then downloaded by clients using HTTP requests over TCP. 
Each HAS client selects the best representation to download (i.e., the best encoding rate and resolution) independently from the other clients. Therefore HAS systems are able to respond to the heterogeneous demands of several HAS clients in a fully distributed and adaptive way.  
The bitrate to download is usually selected by taking into account both the download rate of the previous chunks and the status of the playout buffer, with the aim of maximizing the downloaded bitrate while minimizing the possibility of rebuffering events. 

One of the most challenging aspects in HAS systems is the proper design of the adaptation logic (i.e., the selection of the bitrate to request) at the client side. An intense research has focused on designing HAS client controllers that guarantee a stable and fair utilization of the network resources among multiple clients sharing the same bottleneck.  However, most of this research aims at reaching rate fairness among clients rather than quality fairness. 
Ideally, video distribution solutions should share the bandwidth in such a way that the different users experience a similar video quality. Unfortunately, since video sequences generally have different characteristics, equal rate allocation among clients (rate fairness) does not necessarily translate into quality fairness. From this point of view,  the complete freedom left to  HAS clients  that selfishly  maximizes their own download bitrate  reveals its drawback. To overcome this main limitation,  the  MPEG group is developing an extension of the DASH standard called Server and Network Assisted DASH (SAND) \cite{sand}. SAND is based on asynchronous client-to-network and network-to-network transmissions aimed at improving the Quality of Service (QoS) without interfering with the delivery of the media stream. In this spirit,  we focus on the bitrate selection problem in order to increase the overall QoS of the clients and therefore improve the quality fairness.  

\begin{figure}[t]
\centering
\includegraphics[scale=0.14]{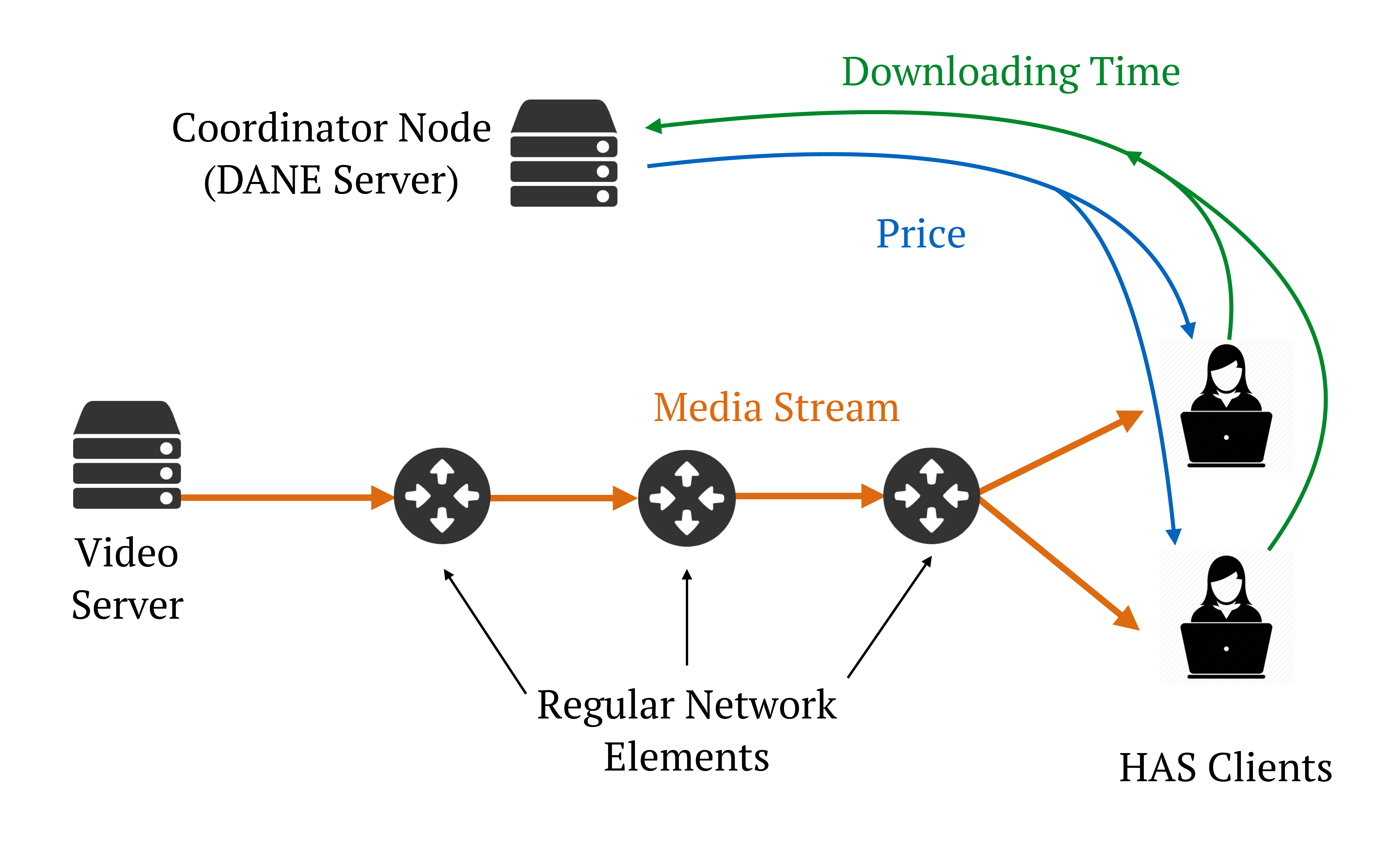}
\caption{Proposed system architecture.}
\label{fig:arch}
\end{figure}

Inspired by  the well known Network Utility Maximization (NUM) framework in congestion controllers~\cite{kelly}, we design a price-based distributed controller, that  maximizes the overall delivered QoS and improves the QoS fairness among users  while respecting the guidelines of the SAND extension. More in details, we consider a multi-users HAS system where clients share a common bottleneck. We define an objective function to properly map the encoding rate of the downloaded representations to the  QoS delivered to clients. Typically, different video characteristics lead to different objective functions.  
We then define the  congestion level of the network as a function of the downloading times of the chunks, which value can easily be measured by the clients. We introduce a coordination node, which corresponds, for example, to a DASH-Assisting Network Element (DANE) in the SAND terminology. As shown  in Fig.~\ref{fig:arch}, this node does not have to lie on the media delivery path, which facilitates the deployability of the proposed solution. The coordination node gathers the the downloading times of the chunks from the HAS clients and  iteratively updates the price value accordingly, the updated price is then sent back to the clients. By following an appropriate price-based bitrate selection policy,  users with simple video sequences, i.e., low bandwidth requirements, do not increase the bitrate of the requested chunks in congested periods in favor of users downloading more complex videos. 
This policy ultimately leads to a higher overall QoS  of the HAS system and to a quality-fair resource allocation.  We test the proposed solution in a network simulator (NS3) under different network conditions and we compare it with other rate-fair controllers proposed in the literature. The simulation results confirm that the achieved rate allocation leads to a better quality fairness among the users with respect to the baseline rate-fair HAS controllers. Moreover, we show the ability of our new algorithm to coexist with TCP cross-traffic and other HAS controllers. 

In summary, the main contributions of the paper are the following: $i)$ we propose a distributed HAS controller that targets quality fairness among several HAS clients sharing a common bottleneck; $ii)$ we introduce a method for measuring the congestion level of a bottleneck link for HAS that relies exclusively on client measurements; $iii)$ we design our controller such that it can be integrated in the SAND architecture; $iv)$ we carry out performance simulations with a realistic network simulator that shows the benefits of the proposed solution. 
 
The paper is structured as follows. 
In Section \ref{related}, we report some related works on the QoS enhancement in HAS systems. In Section \ref{system}, we provide a description of the considered framework. In Section \ref{prob}, we derive the theoretical foundation of the proposed bitrate selection strategy using a simplified model. In Section \ref{impl},  we describe in detail the practical implementation of the controller. We present in Section \ref{results}  the simulations results. Finally, conclusions are provided in Section \ref{conc}.

\section{Related Works} \label{related}
Since a complete description of the  whole literature in adaptation algorithms for HAS would not possible due to space limitations, with the following we discuss  the works that focus on quality-fairness in HAS.

In~\cite{new_schroeder}, the authors optimize the bitrate selection in order to maximize the Quality of Experience (QoE) among a set of HAS users on a wireless link. In this case the base station carries out the optimization according to the different video characteristics. Though this system is able to effectively allocate the available bandwidth it has some drawbacks in terms of deployability, it requires to modify a network element that lies on the delivery path, and scalability, the base station has to collect all the information about the users' videos and solve the optimization problem.
In our proposed system the coordinator is not responsible for solving the optimization problem and it does not need to hold any per user information, thus preserving system scalability.

Several works \cite{SDN,cofano,holland} have proposed solutions for improving DASH QoS based on Software Defined Networking (SDN). The common feature of these solution is the presence of a central network controller that controls the video flows that are currently active in the network. While SDN is a promising technology  to improve Internet performance, it is not currently deployed on a wide scale, therefore solutions based on this technology are not suited for many of the nowadays networks.
In this work, we rather aim at improving the QoS in HAS with an algorithm that exclusively works at the application level and does not assume any particular technology about the inner network nodes. 



In \cite{QL2}, the authors propose a Q-learning multi-agent system for HAS users sharing a common bottleneck in order to maximize a global QoS metric. The problem is formulated as a reinforcement learning problem where the HAS user represents the learning agents. Although this method ultimately achieves the optimal bitrate selection, it requires a very long training phase to learn the optimal solution, making the deployability of this system in realistic environments problematic. In our case we use a model-based formulation therefore we do not require any learning phase and we quickly converge to the optimal bitrate selection.

\section{System Model}\label{system} 
We describe in detail the framework studied in this paper.
We consider a HAS system  with $N$ users, or clients, sharing a bottleneck link with an unknown available capacity $C$. 
This scenario though not general is quite common, think for example about the case where the $N$ users share the same access link or the case where the server access link is the bottleneck. In the event that a heavy traffic load is detected on these links the group of users can ostensibly be gathered.

Each  client downloads video chunks of time duration $T_{ck}$ by sending HTTP requests to the server. The client then stores  the received video data in the  playout buffer, which has a maximum capacity of $M$ chunks.  After a chunk is downloaded the next one is requested immediately if a free slot is available in the buffer, otherwise the client waits until a chunk is played and a slot becomes free to request the next one. When the buffer is full, requests therefore are made every $T_{ck}$ in stationary regime.

Let  $r_i$ be the bitrate of the last chunk downloaded by user $i$, and  $\mathbf{r}=[r_1, r_2, \ldots, r_N]$ be  the  vector corresponding to the bitrates of all the recent clients requests.
We denote by $\tau_i(\mathbf{r})$ the  downloading time for client $i$, defined as the time necessary for user $i$ to download  a chunk encoded at rate $r_i$.
Note that $\tau_i(\mathbf{r})$    depend	s on the entire vector $\mathbf{r}$, since the bottleneck is shared by all users.
We   denote  the rate vector $\mathbf{r}$  as \emph{sustainable} if  $\tau_i(\mathbf{r})\leq T_{ck}, \, \forall i$. A sustainable rate vector implies that users download their chunks in an amount  of time that is sufficient to avoid buffer underflow.

Note that the downloading time $\tau_i(\mathbf{r})$ is an extremely complex function in reality, and represents the network response to the client requests. It depends on the capacity $C$ of the bottleneck link, on the starting time of the downloads, as well as on the random fluctuations of the TCP rate due to packet losses. 
For the sake of simplicity, we first assume an ideal TCP behavior, which means that: $i)$ the bandwidth is always equally  shared among the active connections, $ii)$ the channel is fully utilized when at least one connection is active. Note that these are the ideal characteristic of every rate-fair congestion control algorithms. We then use a realistic TCP connection to evaluate our controller in the conducted experiments.

We define $U_i(r_i)$ to be a strictly increasing concave utility function that represents the quality experienced by user $i$ when the video is downloaded at bitrate $r_i$. Utility functions of different users have different shapes to model the different bandwidth requirements for different video sequences. We finally define the overall QoS of the system as the sum of the single utility functions experienced by each user, more formally,  $\mathcal{U}(\mathbf{r})=\sum_{i=1}^N U_i(r_i)$.

%

\section{Quality-Fair HAS Congestion Controller}\label{prob}
In this section we derive the theoretical foundation of the proposed  controller. 
We focus on the bitrate selection of the users at regime, which means that users need to experience a stationary average downloading time smaller than or equal to $T_{ck}$, in order to avoid buffer underruns, and they request one video chunk every $T_{ck}$.
Rebuffering phases and proper buffer management policies are considered later in the practical implementation of the controller, which is described in the next section. 

We formulate a utility maximization problem for the multi-user system at regime.
The goal is to find a rate vector $\mathbf{r}$ that is sustainable and that maximizes the aggregate utility. This can be achieved by solving the classical NUM problem:
\begin{equation}
\begin{aligned}
& \underset{\mathbf{r}}{\text{maximize}}
& & \sum_{i=1}^{N} U_{i}(r_{i})\\
& \text{subject to}
& &\sum_{i=1}^{N} r_{i} \leq C.
\label{NUM}
\end{aligned}
\end{equation}
The problem consists in maximizing a concave objective function of utilities subject to a linear inequality constraint on the cumulative bitrate.

The optimization problem in \eqref{NUM} can be solved using a dual algorithm, see~\cite{kelly,tutorial}.
The Lagrangian of the problem in \eqref{NUM} corresponds to:
\begin{equation}
L(\mathbf{r},{\mu})= \sum_{i=1}^{N} U_i(r_{i}) + {\lambda} \left( \sum_{i=1}^{N} r_{i} - C \right),
\label{lag}
\end{equation} 
where $\lambda$ is the dual variable, or price, associated to the bottleneck capacity constraint. The optimal solution of the problem can be determined by solving iteratively the following  system of discrete dynamic equations:
\begin{subequations} 
\begin{align}
\label{solveex_a}
   {r}^{k+1}_i &= \left[{U'}_i({\lambda^k})\right]^{-1}\ \  i=1...N\\
\label{solveex_b}
  {\lambda^{k+1}}&=  \Bigg( \lambda^k + \beta  \left( \sum_{i=1}^{N} r^{k+1}_{i} - C \right) \Bigg)_+
\end{align}
\label{solveex}
\end{subequations}
where $\left[{U'}_i(\cdot)\right]^{-1}$ represents the inverse of the derivative of the utility function of user $i$, $()_+$ denotes the projection onto the positive orthant and $\beta$ is a simple parameter to set the speed of change of the dual variable.
Note that users can compute the first step, Eq.~\eqref{solveex_a}, independently, if they know the value of the dual variable $\lambda$.
For evaluating the second step, Eq. \eqref{solveex_b}, the value of the capacity $C$ needs to be known. However this quantity cannot be determined handily since its value depends on protocols overheads and potential cross traffic (which cannot be known in advance). 

We need therefore to modify the second step of the iterative algorithm in order to avoid the explicit the value of the capacity $C$. We thus propose to use the maximum downloading time $\tau_{MAX}=\max_{i=1...N}\tau_{i}$ in place of the rate sum.
According to the ideal TCP behavior described in the previous section, when  $\sum_{i=1}^{N} r_{i} \leq C$ the rate vector $\mathbf{r}$ is sustainable since the total amount of data can be downloaded in less than $T_{ck}$, similarly when  $\sum_{i=1}^{N} r_{i} > C$ the rate vector  is not sustainable\footnote{This follows directly from the assumption that when a single TCP connection is active the channel is fully utilized, and that users request at least one chunk every $T_{ck}$ to avoid buffer underruns.}.
We can therefore map the sum rate constraint into a downloading time constraint, leading to the following equivalent conditions:
\begin{equation}
\sum_{i=1}^{N} r_{i} \leq C \iff \tau_{MAX}(\mathbf{r}) \leq T_{ck}
\label{eq:equiv}
\end{equation}

By using the above equivalency we modify the dynamic system in \eqref{solveex} as follows:
\begin{subequations} 
\begin{align}
\label{eq:final_a} {r}^{k+1}_i &= \left[{U'}_i({\lambda^k})\right]^{-1}\ \  i=1...N\\
\label{eq:final_b} {\lambda^{k+1}}&= \left( \lambda^k + \beta (  \tau_{MAX}(\mathbf{r}^{k+1}) - T_{ck} ) \right)_+.
\end{align}
\label{eq:final}
\end{subequations}
The first step has not changed, but the second step of Eq. \eqref{eq:final_b} can now be easily computed since every user knows the downloading time of the requested chunks, and the maximum value can easily be extracted. The capacity value is not used explicitly anymore, however it is implicitly included in the downloading time measurement $\tau_{MAX}(\mathbf{r})$.
Since the constraints in \eqref{eq:equiv} are equivalent, \eqref{solveex} and \eqref{eq:final} converge at equilibrium to the same rate vector $\mathbf{r}$.

We give now a brief discussion of how the iterative steps of system \eqref{eq:final} can be computed in reality. The adaptation logic, i.e., the selection of the bitrate at the client side is represented by Eq.~\eqref{eq:final_a}, while the price update of the coordinator node is given by Eq.~\eqref{eq:final_b}.
In more details, in the first step, Eq.  \eqref{eq:final_a}, all the users independently compute the optimal bitrate and request the chunks to download at the next iteration accordingly. After the download every user sends to the coordinator node the measured downloading time. The coordinator then performs a maximum pooling operation on the received downloading times and updates the dual variable $\lambda$ using Eq. \eqref{eq:final_b}. The value of $\lambda$ is then sent to the users for the next bitrate selection. By performing these steps iteratively, the system converges to the optimal equilibrium point.

The iterative solution in \eqref{eq:final} represents a modification of the solution of classical NUM problems for the case of HAS system. By using the downloading time of the chunks we can detect an overuse of the available bandwidth without requiring the knowledge of its actual value.
Finally, note that the equivalency of the two conditions in Eq.~\ref{eq:equiv} is true only if the ideal characteristic of the congestion control is verified. If this assumption does not hold, the equivalency is only an approximation whose accuracy depends on the actual behavior of the congestion control. As a result in the real world we need to consider the usage of the downloading time condition instead of the original rate condition as an heuristic approximation suggested by ideal assumption on the congestion control used. Nevertheless the update rules in \eqref{eq:final} are extremely important as to derive a cooperative adaptation strategy for HAS users.


\section{Controller Implementation}\label{impl}
\begin{figure}
\centering
\includegraphics[scale=0.22]{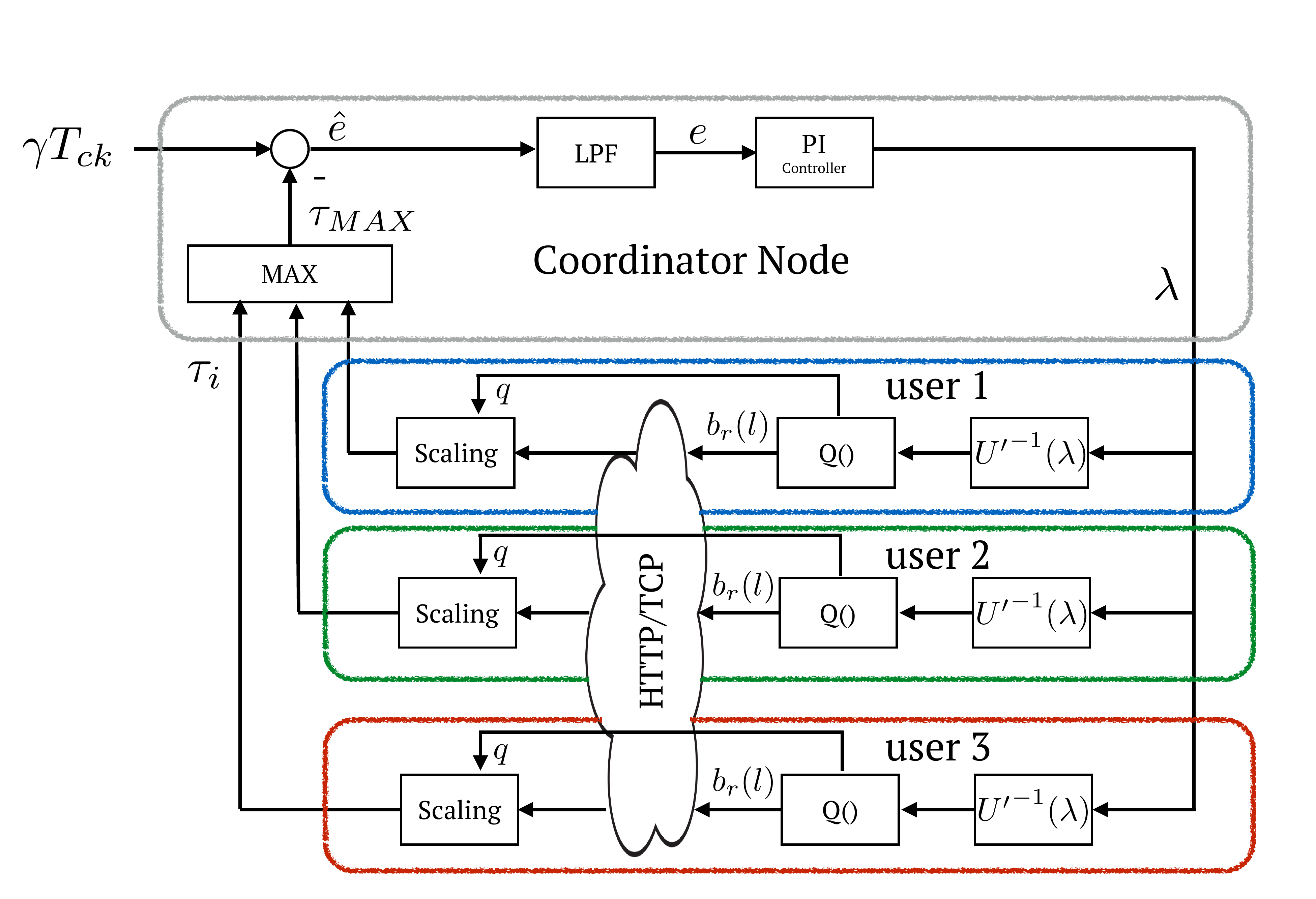}
\caption{Simplified block diagram of the overall system.}
\label{fig:block_dia}
\end{figure}

From the theoretical study of the previous section, we now show how to adjust the iterative solution in~\eqref{eq:final} for it to be used in HAS system in practice.
In particular we consider a discrete rather than continuous set of bitrates, as well as the actual playout buffer management.
The overall HAS multi-user system, depicted in Fig.~\ref{fig:block_dia}, can be seen as a control loop composed of two main entities: the coordinator node, which receives the downloading time measurements from the users and updates the price $\lambda$ accordingly; and the users, which receive the price from the coordinator node and perform the chunk requests based on the video characteristic and the updated price.

\vspace{2mm}

\subsection{Coordinator Node}
In Algorithm~\ref{alg1}, we present the operations that are executed by the coordinator node to update the price $\lambda$  at every iteration step.   The key point is to have a coordinator node that stays as simple as possible  without any need for a per user state information, such that the scalability of the system is preserved. 

Since users are not synchronized the coordinator node processes one user transmission per time. For each downloading time measurement $\tau_{i}$, received from client $i$, the coordinator updates the current maximum downloading time (line 2) and returns the last updated value of the price to user $i$. 

The second part of the algorithm is executed every $T_{ck}$ and corresponds to the price update. 
In order to compute the error signal, $\hat{e}$, the coordinator node needs a reference signal $\gamma T_{ck}$, which expresses the value of the maximum downloading time that the users must have at equilibrium, with $\gamma\in [0,1]$ being a multiplicative factor\footnote{Ideally the value of $\gamma$ should be set to $1$ to fully utilize  the channel. In practical systems, however, we  observed that   $\gamma=0.9-0.95$  provides less noisy results at the cost of marginal channel under-utilization.}.
The error between the maximum downloading time   and the reference downloading time is evaluated (line 6) and  filtered by a Low Pass Filter (LPF), implemented as an Exponential Weighted Moving Average (EWMA) with a coefficient equal to $\alpha_e$, (line 7). The  filter is necessary since  the maximum downloading time is a noisy measure in realistic settings, due to the random behavior of multiple coexisting TCP flows. 
Finally,  the error is integrated according to Eq.~\eqref{eq:final_b} (line 8).
Since the complete control loop is composed also by non-linear blocks, e.g., the utility functions, we limit the value of the integral error to zero in order to avoid integral windup effects \cite{windup}. 

The value of the final price $\lambda$ is then calculated by combining the integral error and the proportional error (line 9), where  $K_P$ and $K_I$ represent the proportional and integral gain respectively. Compared to Eq.~\eqref{eq:final_b} we add in the practical implementation a proportional error to improve the stability of the system without affecting the equilibrium point. The values of $K_P$ and $K_I$ must be set in order to guarantee the stability of the system, i.e., ensuring that the loop return ratio of the control loop has a positive phase margin at the cross frequency. 
Note that $\lambda$, as in Eq.~\eqref{eq:final_b}, is restricted to be positive since negative prices have no meaning.


\begin{algorithm}[t]                      
\caption{{Coordinator Algorithm}}          
\label{alg1}   
\begin{algorithmic}[1]
\If {New downloading time received from user
}
\State $\tau_{MAX}\gets \max(\tau_{MAX},\tau_{i})$
\State {Send most recent $\lambda$ to the user}
\EndIf
\State
\Loop \Comment {executed every $T_{ck}$}
\State {$\hat{e}:=\tau_{MAX}-\gamma T_{ck}$}
\State {$e\gets \alpha_e e + (1-\alpha_e) \hat{e}$}  \Comment {LPF} 
\State {$e_I\gets\max(0,e_I+{e})$}
\State {$\lambda \gets \max(0, K_P {e} + K_I e_I)$} \Comment{Update price}
\State {$\tau_{MAX}\gets 0$} \Comment{Reset $\tau_{MAX}$}
\EndLoop
\end{algorithmic}
\end{algorithm}

\begin{algorithm}[t]                     
\caption{Client Controller Algorithm}          
\label{alg2}   
\begin{algorithmic}[1]
\If {Buffer\_full or download active}
\State \Return
\EndIf
\State
\State $r_{coord} := \left[ {U'}(\lambda/\kappa) \right]^{-1}$
\State $\hat{r}_{TCP}:= \text{last chunk TCP throughput}$
\State $\hat{\alpha}_{TCP}:= \alpha_{TCP}(\text{now - last TCP throughput update})/T_{ck}$
\State $r_{TCP} \gets \hat{\alpha}_{TCP}r_{TCP}+(1-\hat{\alpha}_{TCP})\hat{r}_{TCP}$ \Comment{LPF}
\State $r:=r_{coord}$
\If {$(r_{TCP}<r_{coord})$ and $ (B<T_{ck}(0.6 M))$ }
\State $r \gets r_{TCP}$
\EndIf
\State $B := \text{BufferLevel}()$
\State $\delta:=\max\left(1.0,\min\left(0.25,\frac{B}{T_{ck}(0.7 M)}\right)\right)$
\State {$l \gets \arg \max_{b(l')<r \delta} b(l') $}
\If {$l<l_{old}$}
\State {$l \gets \max(l_{old}-1,l_{min})$}
\EndIf
\If {$l>l_{old}$}
\State {$l \gets \min(l_{old}+1,l_{MAX})$}
\EndIf
\State $\hat{\tau}:= \min(\text{last downloading time},1.25 T_{ck})$
\State $\tau \gets \alpha_{\tau}\tau+(1-\alpha_{\tau})\hat{\tau}$ \Comment{LPF}
\State $\hat{q} := \max(1.0,r_{coord,old}/b(l_{old}))$
\State $q \gets \alpha_{q}q+(1-\alpha_{q})\hat{q}$ \Comment{LPF}
\State send the chunk request for bitrate $b(l)$
\State send the corrected downloading time to coordinator $q\tau$
\end{algorithmic}
\end{algorithm}

\vspace{2mm}

\subsection{Client Controller}\label{client_sub}
We now describe the main steps of the HAS client controller. The behavior of the controller is strongly based on Eq.~\eqref{eq:final_a}.  However, we cannot simply use the aforementioned equation  since buffer level variations as well as discrete sets of available  bitrates need to be taken into account in practice.
The full client algorithm, provided in Algorithm~\ref{alg2}, is executed every time a chunk can be downloaded, i.e., anytime a download is finished and the playback buffer is not full (see downloading conditions -- line 1,2). 

As a first step,  the client controller calculates the value of the ideal bitrate $r_{coord}$ from the last received price value $\lambda$ according to Eq.~\eqref{eq:final_a}. The coefficient $\kappa$ is necessary to normalize the value of the price accordingly to the shape of the utility function and to assure the stability of the system.
In a theoretical model, the controller would request  a chunk  of rate $r_{coord}$. However, we cannot fully neglect the experienced   TCP throughput as well as the client buffer status in realistic implementations.  For example, if the  buffer level is very low and the rate suggested by the coordinator system is remarkably high compared to the measured TCP throughput, it might be a good idea to ignore $r_{coord}$ and select the chunk according to the measured bandwidth only. This situation can occur during the startup phase or after a sudden drop of the available bandwidth.
Therefore, the controller estimates the TCP throughput as described in~\cite{panda} (lines 5-7) and  selects which rate to use between the TCP throughput, $r_{TCP}$, and the ideal rate, $r_{coord}$. Basically, it selects the TCP throughput estimation only if $r_{TCP}<r_{coord}$  and if the video buffer level is below a certain threshold (we set this threshold to be equal to $60\%$ of the maximum buffer occupancy since it offers a good tradeoff between avoiding buffer underruns and trusting generally the coordinator price) (lines 9-10).

Next, rather than selecting exactly $r$, the controller will search for a discounted value $r \delta$, with the discount factor $\delta$ defined in line 12, depends on the buffer level occupancy $B$. The discount factor  is usually $1$ at regime, but it reduces during re-buffering phases in order to decrease the rate of the requested chunks and refill the buffer faster. The discount factor takes values between $0.25$ in low buffer conditions, and $1$ when the buffer occupancy is higher than 70\%.
Finally, the controller selects the chunk with the encoded bitrate that is closest  to $r \delta$. In order to select the bitrate level $l$ we first select the maximum bitrate lower than $r\delta$ (line 13) ($b(l)$ is the encoding bitrate for the representation $l$). Secondly, since large quality variations can be badly perceived by the user, we limit the variation of the representation index with respect to the previous selection $l_{old}$ (lines 14-17).
We then consider the downloading time of the previous chunk $\tau_{old}$ and we filter this variable using a LPF implemented as an EWMA with a coefficient equal to $\alpha_{\tau}$ (line 18-19).  The clipping and filtering  of the downloading time is necessary to improve the coexistence with TCP in practice. When users  compete against  TCP flows,  they can   experience  episodic downloading times that are remarkably larger then the average one, which may cause an unjustified price increase.

The last step of this practical implementation takes into account the quantization of the selected chunk rates, which affects   the granularity of the downloading time values. Due to the rate discretization  we cannot always guarantee an  average maximum downloading time that matches exactly  $\gamma T_{ck}$. This can lead the controller to  frequent oscillations in the price that then translate into annoying oscillations in the users bitrates selection.  To overcome this problem,  we introduce a new variable $q$, which keeps track of the ratio between the ideal rate and the actual requested bitrate (line 20-21). The value $\alpha_q$ corresponds to the coefficient of the EWMA, and $r_{coord,old}/b(l_{old})$ is the ratio between the previous ideal request, $r_{coord,old}$, and the previous chunk request, $b(l_{old})$.  The key point is to perform an  upscaling of the measured downloading time based on the experienced quantization step.  In this way we are able to decrease the difference between the average downloading time of the most demanding user and the reference signal $\gamma T_{ck}$, reducing the variations of the price. Note that the main drawback of the  downloading time correction technique is an under-utilization of the channel (at regime $r_{coord,old}\geq b(l_{old})$), which is however balanced with the reduction of the frequent oscillations of the video quality. 
An alternative way to solve the bitrate discretization problem is to select the chunks in such a way that the average bitrate is equal to the coordinator rate. This method might be useful if we consider dynamic video complexity, i.e. dynamic utility functions. This is however beyond the scope of this work.
As last step the controller sends to the video server the request for a chunk of bitrate equal to $b(l)$ and sends to the coordinator the scaled downloading time measurement equal to $q\tau$.  

Finally, note that all the clipping operations implemented in the client algorithm are active exclusively during transitory phases, e.g., rebuffering events, therefore they do not affect the bitrate selection at regime.

\subsection{Summary of the Proposed Controller}
We conclude this section by listing some benefits of the proposed system.
\begin{itemize}
\item The coordinator node is extremely simple as it does not require any per user state information. The coordinator uses measurement  collected from the users to compute a unique global signal that is then sent back to the users. Each user uses then this signal in the bitrate selection in order to increase the overall QoS. The bitrate selection is done in a fully distributed way to meet the HAS paradigm.
\item The algorithm requires every user to send the downloading time of every chunk to the coordinator node and to receive the price. 
However, the size of both measurement and price messages is very small (few bytes). Moreover the communication overhead grows only linearly with both the number of users and  the number of chunks. As a result, the proposed system has a limited overhead also for large multi-user systems.
\item The coordinator node can be located anywhere in the network as long as it is able to communicate with the HAS users that share the bottleneck link. 
\item In case of broken communication link between the client controller and the coordinator node, the client can simply fall back to a classical rate/buffer-based HAS controller. 
Users that are not able to communicate with the coordinator node will be perceived as cross-traffic by the other HAS users, which can still perform the optimal selection strategy.

\end{itemize}


\section{System Evaluation}\label{results}
We now provide simulation results to evaluate the performance of the proposed system. We implement the algorithm described in Section \ref{impl}  in the NS3 network simulator and evaluate it  in different representative scenarios.

\subsection{Experimental Setup}
\begin{figure}
\centering
\includegraphics[scale=0.2]{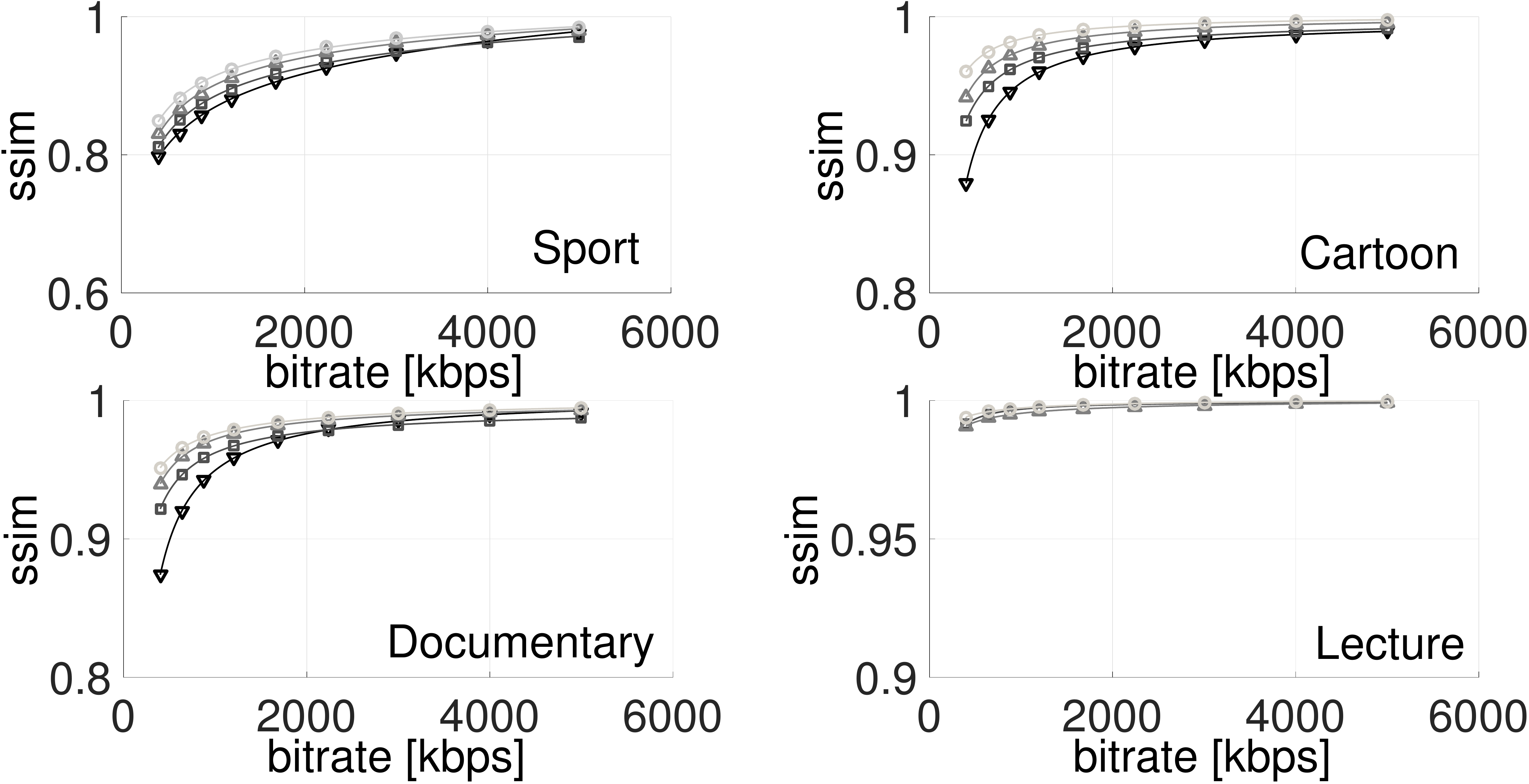}
\includegraphics[scale=0.2]{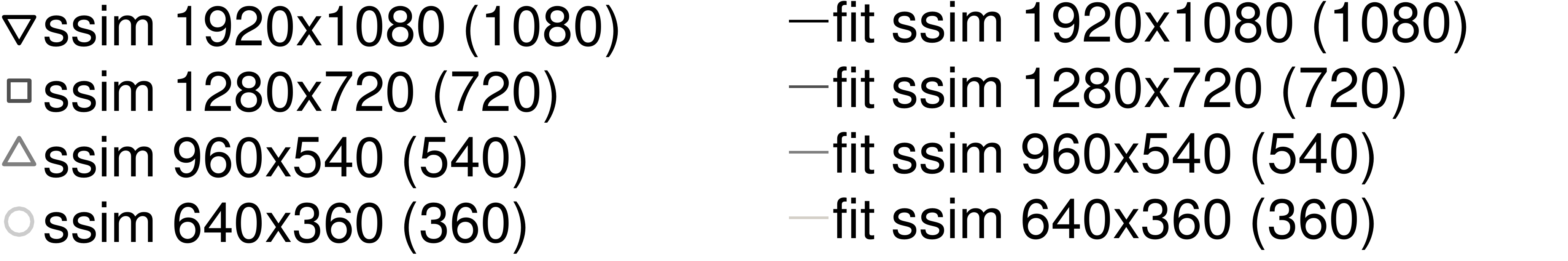}
\caption{Quality-rate utility functions for the video sequences under consideration. Solid lines represent the continuous model of Eq. \eqref{ssim_fit} while symbols are experimental measurements.}
\label{ssimcurves}
\end{figure}

In order to evaluate the proposed algorithm we use the well-known Structural Similarity (SSIM) metric~\cite{ssim} as a utility function.
We consider four types of videos with different properties:  a high motion sport video, two medium complexity videos, a cartoon    and a documentary, and a low complexity lecture video. The original videos have been downscaled to smaller resolutions and every resolution has been encoded at different bitrates using h264 codec~\cite{h264}.
We have then extracted the average SSIM of the encoded sequences at different bitrates. We have derived the following continuous model of the SSIM:
\begin{equation}
U_i(r)=a_i \cdot r^{b_i}+c_i,
\label{ssim_fit}
\end{equation}
where the coefficients $a_i$, $b_i$ and $c_i$ for the encoded sequence $i$ are derived by curve fitting with the SSIM experimental points.
The experimental SSIM data points and the fitting curves  are depicted in Fig.~\ref{ssimcurves}.
Note that a visually pleasant video usually has a SSIM score above $0.8$ and a gain in SSIM of $0.05$ might correspond to an increase of one point in  Mean Opinion Score (MOS)~\cite{choivideo}.

In our simulations we identify each user with a single video at a given resolution, therefore with a single constant utility curve that is then used to execute the adaptation logic described in Subsection~\ref{client_sub}.
We assume that each user  knows the utility function of the  requested video. In reality, this is possible by including this information in the Media Presentation Description (MPD) file of the video, or alternatively, the service provider can  make it available on the server  as secondary information. Another possibility is that the users implement a no-reference distortion model to assess the quality of the displayed video sequence. 

We compare our algorithm with three HAS controllers proposed in the literature, namely a \emph{conventional} HAS controller as described and implemented in~\cite{panda}, the Probe and Adapt (PANDA) algorithm also proposed in~\cite{panda}, and the  ELASTIC algorithm proposed in~\cite{elastic}. These three algorithms  represent well the different behavior that rate-fair controllers can exhibit: PANDA  is more conservative since  it prefers to slightly underutilize the channel at the benefit of having a more constant bitrate selection. ELASTIC, on the other hand, strives to fully utilize the channel at the cost of more frequent quality variations. The conventional controller  offers somehow an average behavior compared to the other two. To have a fair comparison among the different controllers, we fix the maximum buffer size of all the  algorithms to $M$ chunks, and we modify accordingly the parameters that control the buffer size in the baseline algorithms. In particular, the parameters $B_{min}$ of PANDA and $q_T$ of ELASTIC are both set to $6T_{ck}$. The other parameters of the baseline algorithms are set accordingly to the cited works.
Note that in our work we do not consider freezing events as metric of comparison, therefore reducing the size of the buffer does not penalize any of the algorithms. For the proposed quality-fair algorithm, the value of the parameters are listed in Table~\ref{tab:param}. We set the values of these parameters in order to have: $i)$ a good reactivity, thus good speed to convergence, $ii)$ and clean signals,  thus reducing the noise introduced by the network measurements.

\begin{figure}
\centering
\includegraphics[scale=0.125]{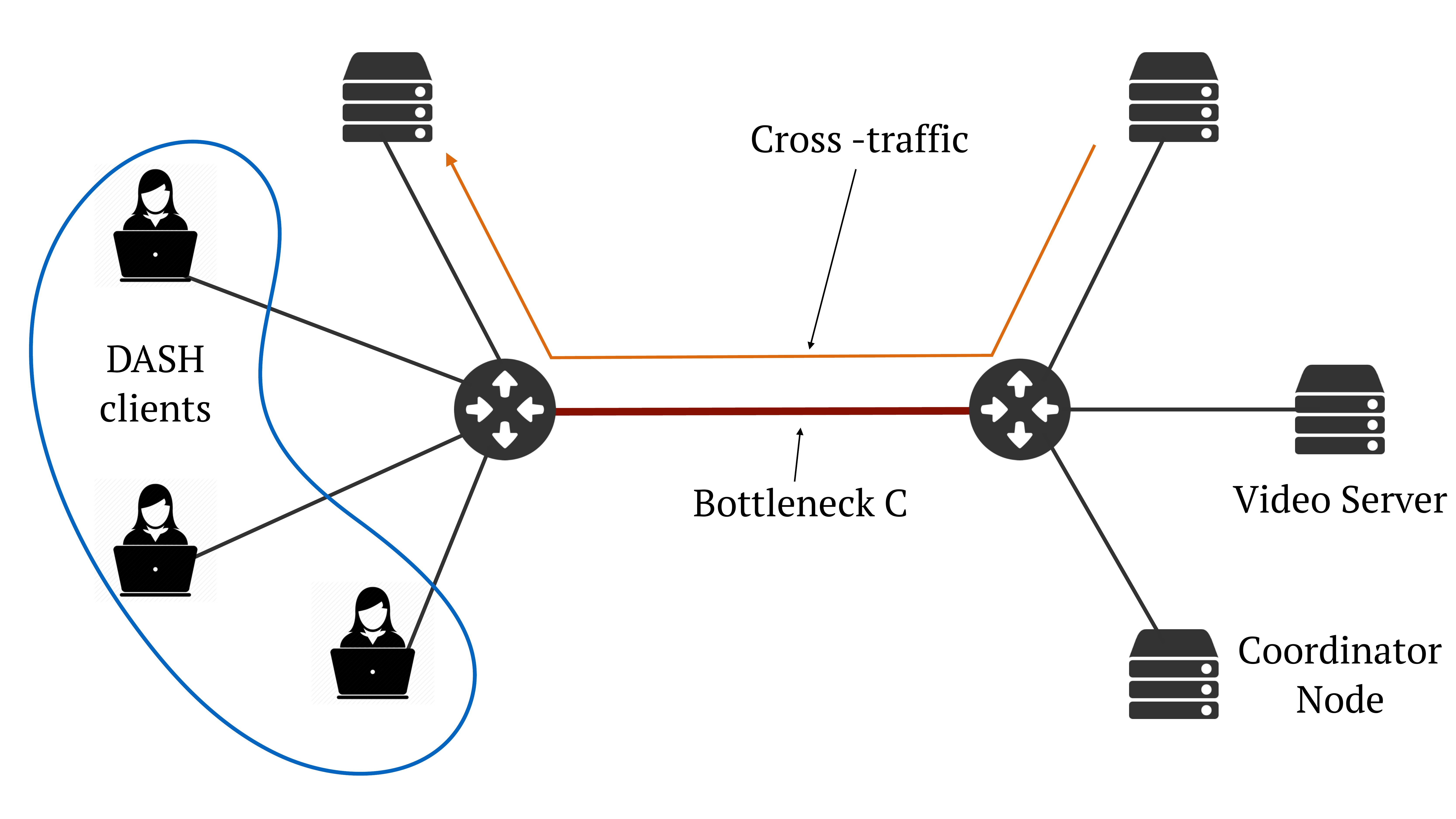}
\caption{Topology used in the different simulated scenarios.}
\label{topo}
\end{figure}

Finally, the proposed controller as well as the baseline algorithms are tested over the network  topology  depicted in Fig.~\ref{topo}, where all users share the same bottleneck link. The links that connect the HAS users to the bottleneck link are local high-speed links.
Lastly, the cross-traffic, if present, shares only the bottleneck link with the other HAS users.


\begin{table}[h]
\centering
\caption{Parameters used in the implementation}
\label{tab:param}
\begin{tabular}{|c||c|}
\hline
Parameter          & value     \\ \hline \hline
\multicolumn{2}{|c|}{All algorithms} \\ \hline 
$T_{ck}$          &      $2$   s       \\ \hline
$M$ (Max buffer size)                &     $10$        \\ \hline
Bitrates available  & \multicolumn{1}{c|}{\begin{tabular}[c]{@{}c@{}}$[400\ 640\ 880\ 1200\ 1680\ 2240\ $\\ $2800\ 3600\ 4400\ 6000]$ kbps\end{tabular}} \\ \hline    \hline
\multicolumn{2}{|c|}{Proposed algorithm} \\ \hline 
$\gamma$          &      $0.95$       \\ \hline
$\alpha_e$          &     $0.75$        \\ \hline
$K_p$          &      $1$       \\ \hline
$K_i$          &       $0.25$      \\ \hline
$\kappa$          &      $1e6$        \\ \hline 
$\alpha_{TCP}$, $\alpha_q$, $\alpha_{\tau}$          &     $0.75$    \\ \hline \hline
\end{tabular}
\end{table}

\subsection{Simulation Results}
We now provide the simulation results carried out in the settings described above.
We show first in detail how the proposed algorithm behaves. Then, we show the gain of our controller with respect to rate-fair controllers when the bottleneck is shared by many HAS users.
In a last set of simulations we evaluate the performance of the proposed algorithm when competing with cross-traffic.

\begin{figure}
\centering
\includegraphics[scale=0.1925]{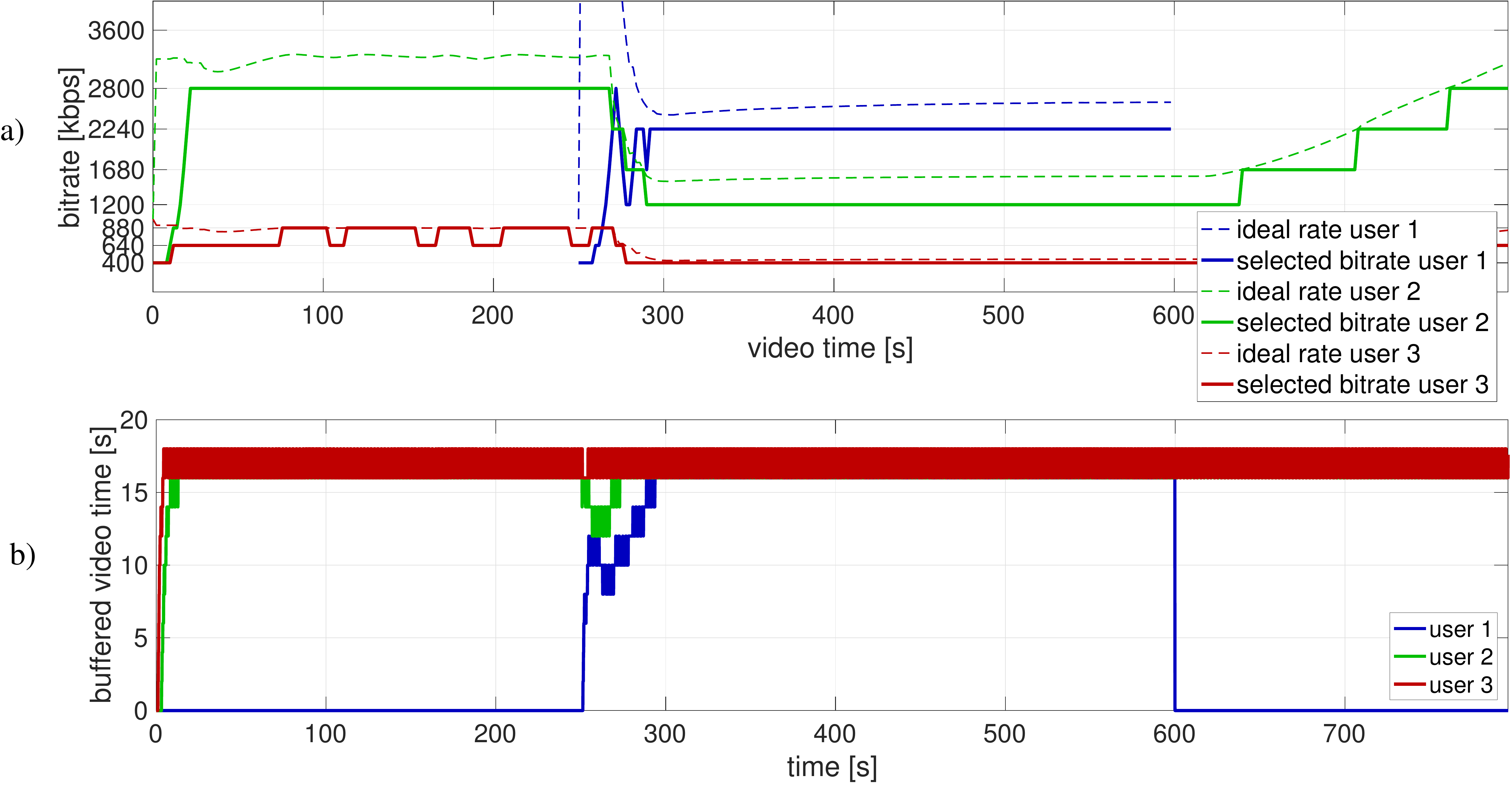}
\includegraphics[scale=0.1925]{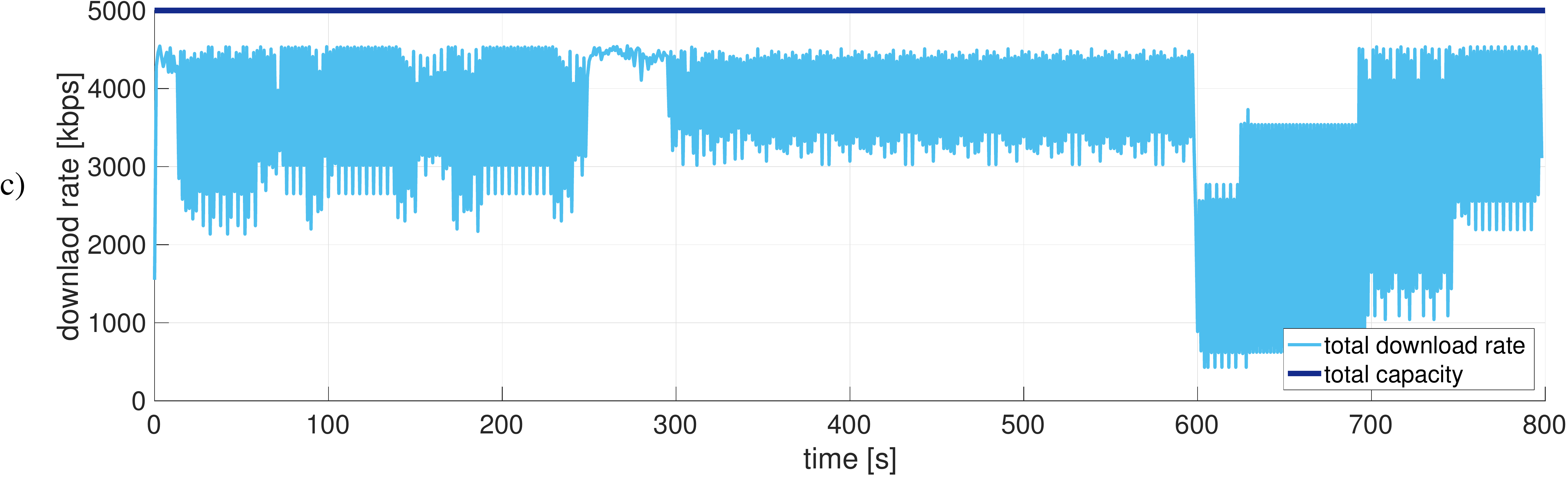}
\caption{Performance of the proposed algorithm when three HAS users implementing our algorithm compete for the same bottleneck channel. The three plots respectively show the selected and ideal bitrates, the buffer occupancy and the channel utilization.}
\label{sim1}
\end{figure}

In the first test case three HAS clients share a common bottleneck link that has a capacity of $5$ Mbps. The Users $2$ and $3$, download the cartoon video at resolution $1080$ and the lecture video at resolution $720$, respectively, from the beginning of the simulation and stay always active, while user $1$ downloads the sport video at resolution $540$, between the timestamps $250$s and $600$s.  
The results are depicted in Fig.~\ref{sim1}. In Fig.~\ref{sim1}a, we provide both the video bitrate selected by the users and the ideal bitrates ($r_{coord}$) as described in Subsection~\ref{client_sub}. This plot shows  the ability of the algorithm to fairly  allocate the available bandwidth when client have different utility functions. Since user $1$ is the one consuming the most complex video sequence, it is also the one that gets a larger portion of the channel link. From the SSIM curves, we know that for a bitrate of about $2.2$ Mbps, user 1 experiences a SSIM value of approximately $0.94$, while user 3 already achieves a SSIM value above $0.98$ at $0.4$ Mbps. Thus, the proposed controller is clearly able to improve the quality fairness among the users with respect to a rate-fair controller, which would allocate approximately $1.5$ Mbps per user, making user 1  suffer of poor video quality while only slightly increasing the quality of user 3.
Fig.~\ref{sim1}b further shows the buffer level of the users. The playout buffers of all the three users have an occupancy level close to the maximum value, and no underruns are experienced during the simulation. The channel utilization, depicted in Fig.~\ref{sim1}c, is also satisfactory. In fact the total download rate, given by the sum of the bitrates requested by all users, settles to a value that is close to the channel capacity. 
Note that the reported channel capacity corresponds to the physical bandwidth, which does not take into account the TCP/IP protocols overhead, thus it is not possible to exactly match its value. 

\begin{figure*}
\centering
\includegraphics[scale=0.3]{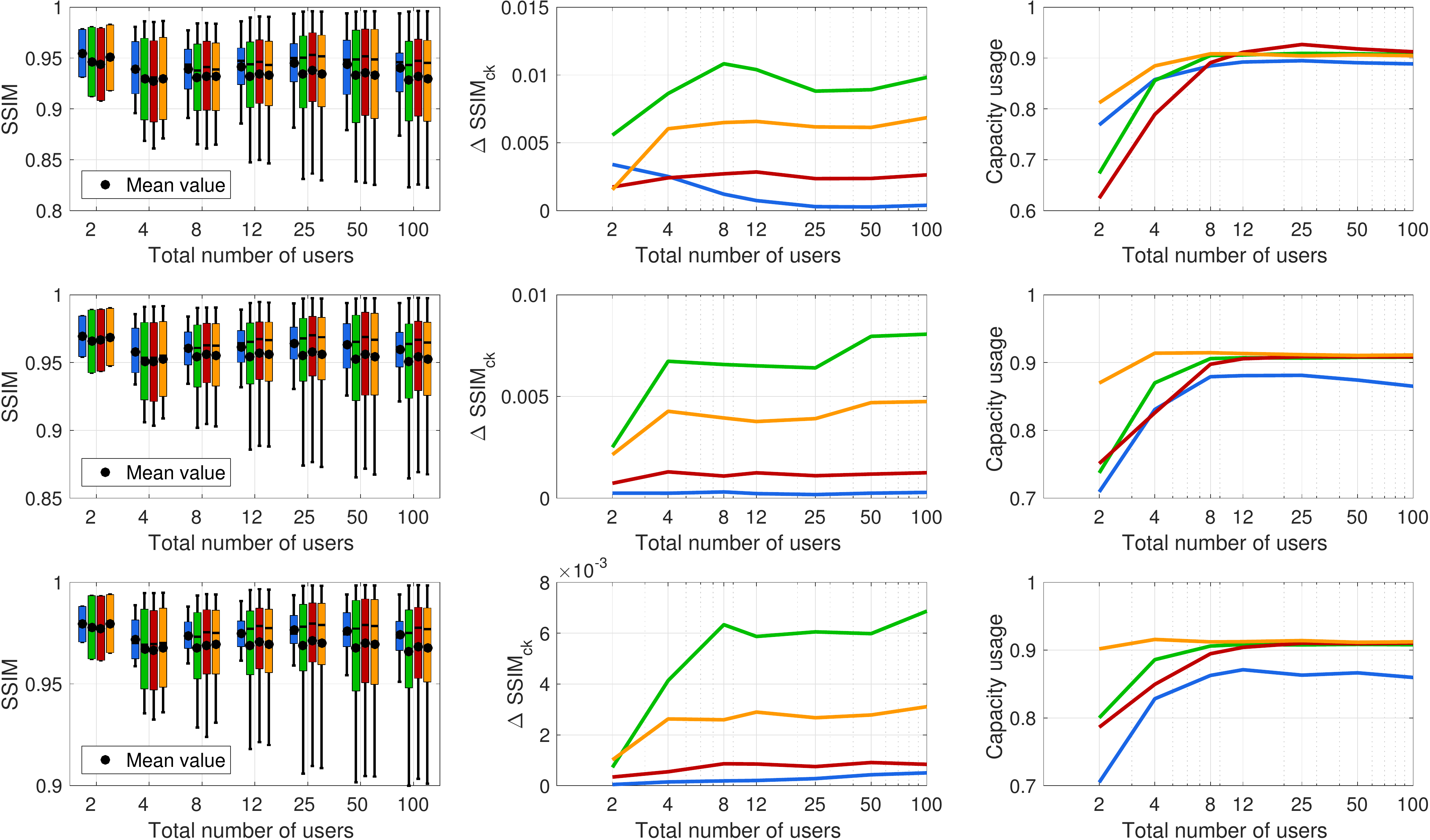}
\includegraphics[scale=0.3]{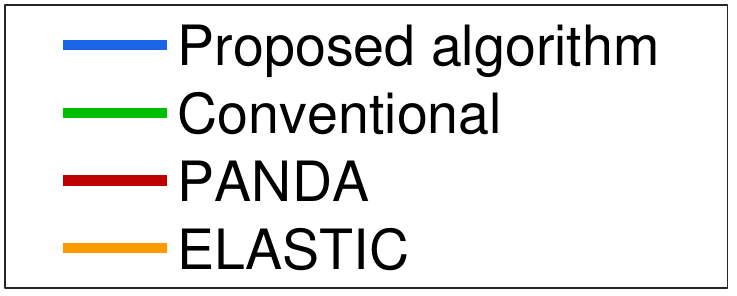}
\caption{ SSIM statistics, SSIM variations and channel utilization for the four implemented controllers for different numbers of users $N$. The per user capacity $C_{usr}$ has been set to $0.75$, $1.25$ and $2.0$ Mbps for the first, second and third row of plots respectively.}
\label{sim_var_usr}
\end{figure*}

We now consider $N$ users, and we randomly assign to each of them a video at a given resolution and thus a corresponding utility curve. We then set    the bottleneck capacity $C$ to $NC_{usr}$, where $C_{usr}$ is  the average  per user capacity. We consider $10$ different realizations of the utility-user random selection, every metric shown in the final plots is the result of the average operation among the  different realization. For each realization, we simulate the video streaming session where all users are simultaneously active for 460 seconds and we evaluate the average SSIM experienced at regime, i.e., after 60 seconds of video. Beyond the  average SSIM, we also compute for each user  the average SSIM variation per downloaded chunk as follows:
\begin{equation}
\Delta\text{SSIM}_{ck}=\frac{1}{L-1}\sum_{l=2}^L |\text{SSIM}(l)-\text{SSIM}(l-1)|,
\end{equation}
where $L$ is the total number of chunks downloaded by the user and $\text{SSIM}(l)$ is the SSIM value for chunk $l$.
After we compute the average SSIM variation for every user, we average this value among the user population of the simulation.
This metric quantifies the average variation of quality level among consecutive chunks and captures possible SSIM oscillations rather than the simple heterogeneity of the SSIM over the all video sequence.  Since it has been shown that frequent quality switches result in QoE degradation~\cite{switchdash}, the lower the $\Delta\text{SSIM}_{ck}$ the better the QoE.  The last metric that we compute is the capacity usage, which is the time average cumulative downloaded bitrate of the users divided by the total capacity. A capacity usage close to $1$ means an efficient use of the available resources. 
The three metrics above are evaluated in scenarios with   different numbers  of users, i.e., $N=[2\ 4\ 8\ 12\ 25\ 50\ 100]$, and   different   per user capacities, i.e.,  $C_{usr}=[0.75\ 1.25\ 2.0]$ Mbps.
The corresponding results are depicted in Fig.~\ref{sim_var_usr}. Every element of the box-plot is composed of $i)$ a rectangle, which represents the first and third quartile divided by the median value $ii)$ the whiskers, which delimit the minimum and the maximum value of the time-average SSIM among the user population and $iii)$ the black dot which corresponds to the mean value over the population. 
We can   notice  that our algorithm is in general able to achieve better average quality compared with the rate-fair controllers. In particular the proposed algorithm is able to allocate more rate to the users that are watching high demanding videos. The minimum average SSIM of the proposed algorithm is  remarkably higher than the one of the rate-fair controllers. By looking at the numerical values, it can be seen that our method can achieve a gain up to $0.05$ points of SSIM for large values of $N$, and a gain of around $0.01$ points of SSIM for small values of $N$. In general, the SSIM gain is larger for larger value of  $C_{usr}$, since there is a larger margin of optimization in this case thanks to the larger amount of total bandwidth that can be re-allocated among the users.
It is also worth noting that all the baseline algorithms show comparable performance among each other since they all target a rate-fair allocation.  
Beyond increasing the average SSIM, the proposed algorithm also reduced the  average SSIM variations. As it is shown  in the second column of Fig.~\ref{sim_var_usr}, this value is substantially smaller than the variations experienced by   the rate-fair controllers.   The PANDA algorithm, since it is the most conservative, is the one behaving best among the three controllers used for comparison, as expected.
From the third column of Fig~\ref{sim_var_usr}, we can notice that the proposed algorithm is the one achieving the lowest bandwidth utilization. Nevertheless, the efficient usage of the bandwidth permits to the proposed algorithm to have better performances in the other metrics. The low bandwidth utilization is caused by the policy of selecting always a bitrate that is lower than the ideal bitrate. By applying a selection policy that targets a bitrate selection that is on average equal to the ideal rate the capacity usage can be increased, at the cost of more quality variations.
Finally, note that we vary the number of users from a simple 2 users scenario to a scenario with 100 users, the proposed algorithm always achieves a better quality fairness with respect to rate-fair controllers, showing that our system scales well to large population of clients.

\begin{figure*}
\centering
\includegraphics[scale=0.3]{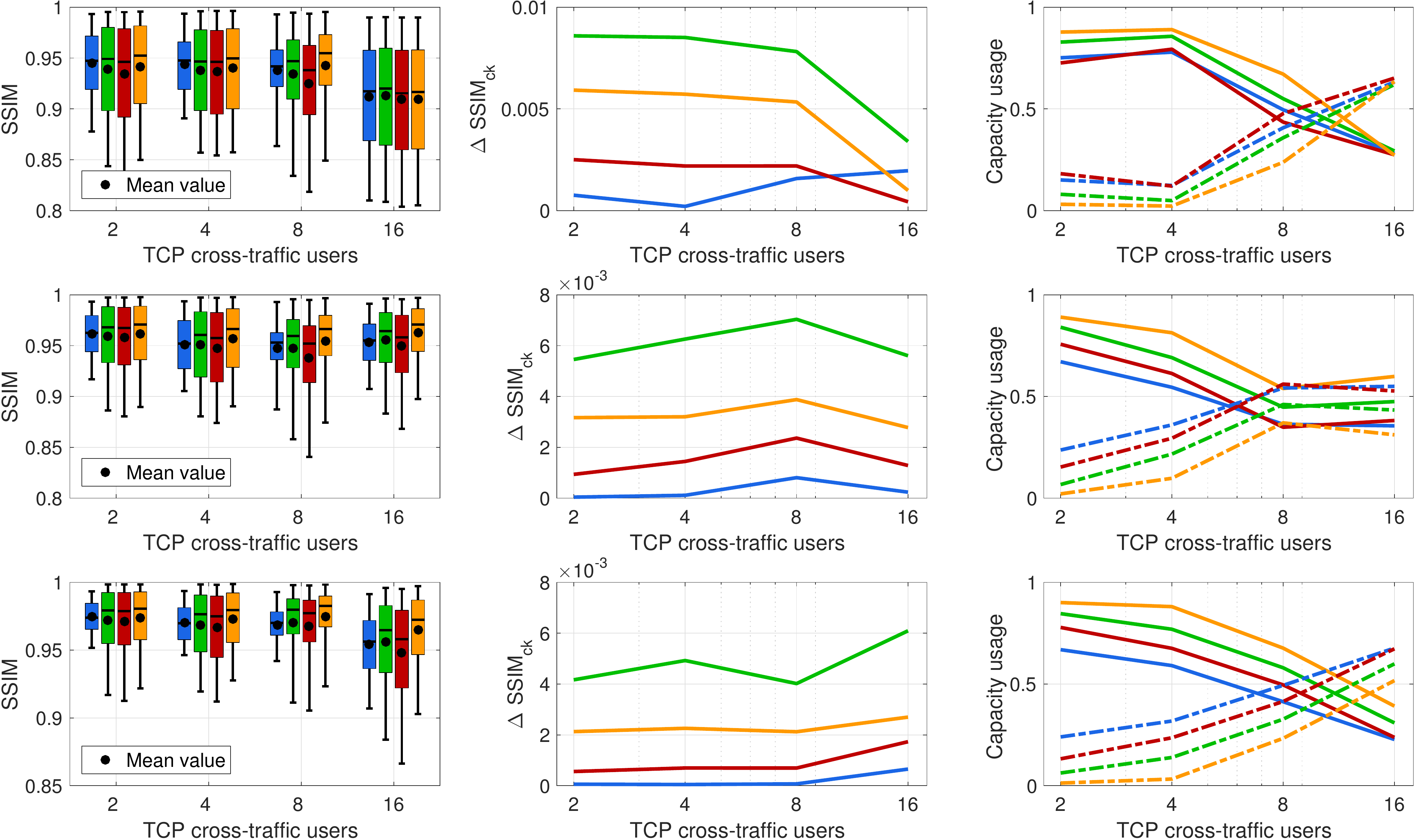}
\includegraphics[scale=0.3]{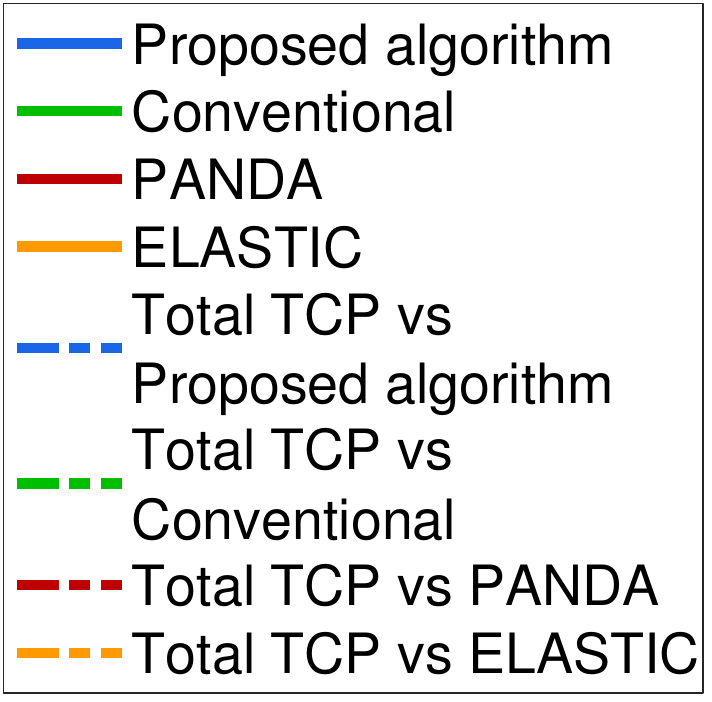}
\caption{ SSIM statistics, SSIM variations and channel utilization for the four implemented controllers for a set $16$ HAS users sharing the bottleneck with a varying number of TCP flows . The per user capacity $C_{usr}$ has been set to $0.75$, $1.25$ and $2.0$ Mbps for the first, second and third row of plots respectively.}
\label{sim_var_usr_TCP}
\end{figure*}

We further analyze the performance of our algorithm when the bottleneck capacity is shared with TCP cross-traffic for different amounts of TCP connections. 
We set the number of HAS users to $N=16$ and then add  different numbers of TCP connections, i.e., $N_{TCP}=[2\ 4\ 8\ 16]$; in percentage the amount of TCP cross-traffic varies accordingly from $11\%$ to $50\%$ of the total connections. We also vary the amount of the total capacity: $C=(N+N_{TCP})C_{usr}$, and the per user capacity is set to $C_{usr}=[0.75\ 1.25\ 2.0] $ Mbps in different simulations.  We then compute the same metrics of the previous tests and the results are shown in Fig.~\ref{sim_var_usr_TCP}.
The average SSIM shows that the different algorithms are able to achieve approximatively the same performance. However, the proposed algorithm achieves higher values of minimum SSIM with respect to the rate-fair controllers. From the second column in Fig. \ref{sim_var_usr_TCP}, we see that the proposed method achieves the lowest  SSIM variations in most of the cases, confirming the behavior  of Fig.~\ref{sim_var_usr}. In terms of channel utilization, ELASTIC is the algorithm that achieves the highest utilization ratio. Our algorithm instead has the lowest channel utilization together with the PANDA algorithm. We further notice that the sum of the HAS users utilization plus TCP utilization (in dashed lines) is close to one, as expected. We finally point out that our algorithm achieves approximatively the same average quality as the other algorithms using less bandwidth. 

\begin{figure}[h]
\centering
\includegraphics[scale=0.18]{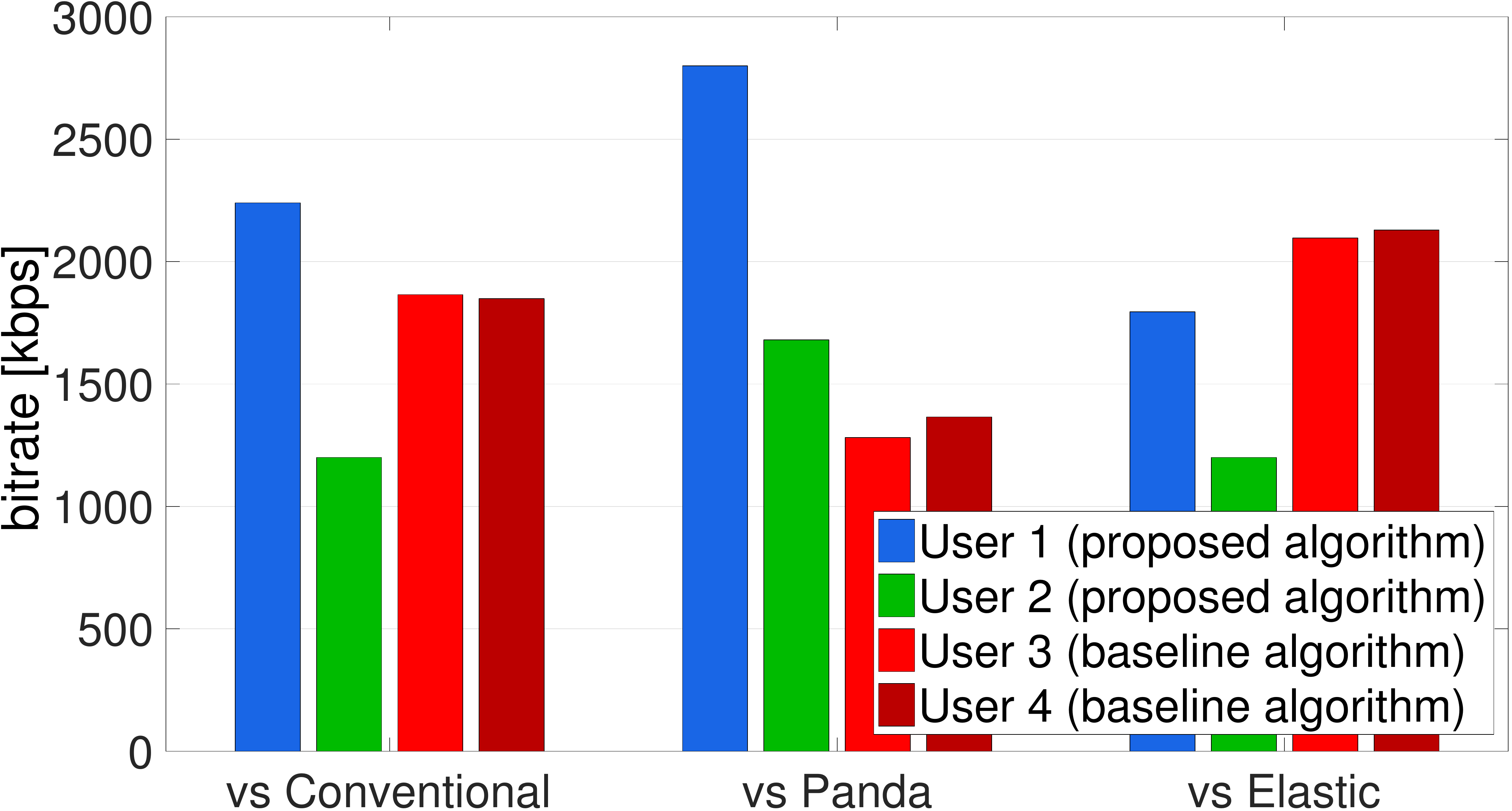}
\caption{Average equilibrium bitrate achieved by the proposed controller when competing with the other rate-fair HAS controllers.}
\label{sim5}
\end{figure}

Finally, in the last set of simulations, we consider the scenario where  only HAS users share the bottleneck channels, but with different controllers  implemented at the client side.  
More in details,  we have $4$ HAS users, two with the proposed algorithm, two with one of the other baseline controllers. The users $1$ and $2$, which implement the proposed algorithm, download a high complexity video and a low complexity one respectively. The baseline controllers (users $3$ and $4$) are content agnostic, thus their behavior does not depend on the utility curve of the videos. The bottleneck capacity is set to $8$Mbps, and all the   users are simultaneously active during the simulation. The results are shown in Fig.~\ref{sim5}. The green and blue bars correspond to the average bitrate requested by  the clients implementing the proposed algorithm, while the two red bars correspond to the average bitrate requested by clients implementing one of the other controllers. The least fair scenario is the one in which the proposed algorithm competes with PANDA. This is expected since, as we have observed in the previous results PANDA is a very conservative algorithm. On the other hand ELASTIC, which is the most aggressive controller, achieves a larger downloading rate when competing with the proposed controller.
The goal of this final tests is to show that the rate-fair HAS controllers neither dominate,nor are dominated by the proposed algorithm and that they can effectively coexist. Consequently we expect that in a scenario with a large number of rate-fair HAS controllers  the performance achieved by our controller are comparable to the TCP cross-traffic results of  Fig.~\ref{sim_var_usr_TCP}.

\section{Conclusions}\label{conc}
In this paper,  we  have proposed a  price-based HAS controller  that is able to enhance the overall QoS and improve quality fairness among HAS clients sharing a common bottleneck link. Based on the experienced downloading times, a  coordinator node  evaluates  the bottleneck price that reflects the congestion level of the network. The users then perform a quality-fair bitrate selection based on this price information. The ideal controller is adapted to work in realistic settings and tested in the network simulator NS3. The  proposed algorithm is extremely scalable in terms of both computation and communication requirements. 
The simulation  results show the ability of the proposed algorithm to work under different network conditions, and to improve the quality fairness of the users when compared to classical rate-fair controllers. The proposed controller is also able to work properly in scenarios where the bottleneck link is shared with TCP and other HAS cross-traffic.
 As future work,  we plan to extend the proposed algorithm to multiple bottlenecks scenarios and to the case of dynamic utility functions, e.g., time varying video complexity.

\bibliographystyle{IEEEtran}
\bibliography{draft.bib}  
\end{document}